\documentstyle[12pt]{article}
\begin{document}

\begin{flushright}
\begin{tabular}{l}
CU-TP-718 \\
UPR-716-T \\
hep-th/9609100 \\
September, 1996 
\end{tabular}
\end{flushright}

\vspace{2mm}

\begin{center}

{\Large \bf 
Vacuum decay and internal symmetries
}
\\ 
\vspace{8mm}

\def\thefootnote{\fnsymbol{footnote}}
\setcounter{footnote}{0}

Alexander Kusenko\footnote{ electronic mail:
sasha@langacker.hep.upenn.edu; address after October 1, 1996:
Theory Division, CERN, CH-1211 Geneva 23, Switzerland
}$^{1}$,  
Kimyeong Lee\footnote{ electronic mail: klee@phys.columbia.edu}$^{2}$
\\ and \\
Erick J. Weinberg\footnote{electronic mail: ejw@phys.columbia.edu
}$^{2}$ 
\\ \vspace{3mm}
$^1$Department of Physics and Astronomy, University of Pennsylvania \\ 
Philadelphia, PA 19104-6396 \\
$^2$Department of Physics, Columbia University, New York, NY 10027

\vspace{12mm}

\end{center}
\centerline{\bf ABSTRACT}
\vskip 5mm
\begin{quote}
{\baselineskip 16pt

We study the effects of internal symmetries on the decay by bubble
nucleation of a metastable false vacuum.  The zero modes about the
bounce solution that are associated with the breaking of continuous
internal symmetries result in an enhancement of the tunneling rate
into vacua in which  some of the symmetries of the initial state
are spontaneously broken.  We develop a general formalism for
evaluating the effects of these zero modes on the bubble nucleation
rate in both flat and curved space-times.

}
\end{quote}

\vfill

\pagestyle{empty}

\pagebreak

\pagestyle{plain}
\pagenumbering{arabic}
\renewcommand{\thefootnote}{\arabic{footnote}}
\setcounter{footnote}{0}

\makeatletter
\renewcommand\theequation{\thesection.\arabic{equation}}
\@addtoreset{equation}{section}
\makeatother

\section{Introduction}  

During its early history the universe may have undergone a number of
phase transitions.  One or more of these may have been first-order
transitions in which the universe was for a time trapped in a
metastable ``false vacuum'' state, with the transition proceeding by
the nucleation and expansion of bubble of the stable ``true vacuum''.
A crucial quantity for the development of such a transition is the
bubble nucleation rate per unit volume $\Gamma$.  A semiclassical
procedure, based on a Euclidean ``bounce'' solution, has been
developed for the calculation of $\Gamma$ at zero temperature
\cite{bounce}, and extended to the case of bubble nucleation at
nonzero temperature \cite{highT}.

Recently \cite{ak} it was noted that there can be an enhancement of $\Gamma$ if a
continuous internal symmetry of the false vacuum is completely or partially
broken by the true vacuum.  This enhancement can be understood as a consequence
of the fact that, instead
of a single possible final state, there is a continuous family of degenerate
true vacuum states into which the decay can occur.  More formally, the effect
arises from the existence of additional zero eigenvalues in the spectrum of
small fluctuations about the bounce solution.

The primary focus of Ref.~\cite{ak} was on the case of a broken $U(1)$ symmetry
(see also \cite{ew1}).  While a similar enhancement is expected for larger
symmetry groups, the treatment of the zero modes becomes somewhat more
complicated for the case of a non-Abelian symmetry.   
In this note we develop the formalism needed to deal with the case of an
arbitrary symmetry.  We also discuss some further implications of these
results, including the extension of these results to bubble nucleation in
curved space-time. 

The remainder of this paper is organized as follows.  In Sec.~2 we develop the
general formalism and estimate the magnitude of the enhancement that can be
achieved.  As a concrete example, we apply this formalism to the case of 
$SU(2)$ symmetry in Sec.~3.  In Sec.~4 we discuss the extension to this work to
curved space-time, using the formalism of Coleman and De Luccia
\cite{cdl}.   We show that although the zero mode contribution to the curved
space-time nucleation rate appears at first sight to be rather different
from its flat space-time counterpart, it does in fact give the expected
result in the the limit where gravitational effects are negligible.  
Section~5 contains some brief concluding remarks.

\section{General Formalism}

    We consider a field theory whose fields we assemble into a column vector
$\phi(x)$ with purely real components.  The standard method \cite{bounce} for the
calculation of the quantum mechanical (i.e., zero temperature)
bubble nucleation rate per unit volume, $\Gamma$, is based on the
existence of a ``bounce'' solution $\phi_b(x)$ of the Euclidean field
equations that tends to
the false vacuum value $\phi_f$ at spatial infinity and  
approaches (although not necessarily reaches) the true vacuum value
$\phi_t$ near a point that may be taken to be the origin.  The result
may be written in the form
\begin{equation}
 \Gamma = {1\over \Omega} {I_b \over I_f}  
\end{equation}
Here $I_f$ and $I_b$ are the contributions to the Euclidean path
integral of $e^{-S_E}$ (where $S_E$ is the Euclidean action) from the
homogeneous false vacuum and bounce configurations, respectively, while
the division by $\Omega$, the volume of the four-dimensional Euclidean
space, arises in order to obtain a rate per unit volume.  The
contribution to the path integral from a stationary point
$\bar\phi(x)$ can be evaluated by expanding the field as
\begin{equation}
  \phi(x) = \bar\phi(x) + \sum_j c_j \, \psi_j(x)
\end{equation}
where the $\psi_j(x)$ form a complete set of orthonormal
eigenfunctions of the second variation operator
$S''_E(\bar\phi) \equiv  \delta^2 S_E/ \delta \phi(x) \delta \phi(y)$.   
For $\bar\phi(x)= \phi_f$, this gives, in leading approximation, a
product of Gaussian integrals over the real variables $c_j$ that results in
\begin{equation}
    I_f = e^{-S_E(\phi_f)} \, \left\{ \det [S''(\phi_f)] \right\}^{-1/2}
     \equiv  e^{-S_E(\phi_f)} \, D_f 
\end{equation}

The calculation of the bounce contribution is complicated by the
presence of zero eigenvalues in the spectrum of $S_E''(\phi_b)$.  Four of these
correspond to translation of the bounce in the four-dimensional Euclidean
space.  We will assume that the remainder are all associated with internal
symmetries of the false vacuum that are not symmetries of the bounce.  
These zero modes 
are handled by eliminating the corresponding normal mode coefficients
$c_i$ in favor of an equal number of collective coordinates $z_i$.   
The zero modes about a bounce configuration $\phi_b(x,z)$ can then be
written in the form
\begin{equation}
   \psi_i(x,z) = N_{ij}(z) {\partial \phi_b(x,z) \over \partial z_j} 
\end{equation}
where the $N_{ij}$ satisfy the equation
\begin{equation}
   [(N^\dagger N)^{-1}]_{kl}  = \int d^4x 
   {\partial \phi_b^\dagger (x;z) \over \partial z_l} \,\,  
    {\partial \phi_b(x;z) \over \partial z_k} \equiv 2\pi (M_b)_{kl}
\label{Mdef}
\end{equation}
that follows from the orthonormality of the $\psi_i$.  (The factor of $2\pi$ is
for later convenience.)

The bounce contribution can then be written as a product of two
factors.  The first
\begin{equation}
       I_b^{(1)}= e^{-S_E(\phi_b)}  {1\over 2} \left| \det{}'[S''(\phi_b)]
      \right|^{-1/2} \equiv  e^{-S_E(\phi_b)} \, D_b 
\end{equation}
arises from the integration over the modes with nonzero eigenvalues.
The prime indicates that the functional determinant is to be taken
in the subspace orthogonal to the zero modes, while the factor of $1/2$
arises from the integration over the single negative eigenvalue mode.
The second factor, from integrating over the remaining $n$ variables,
is 
\begin{equation}
    I_b^{(2)} =  (2\pi)^{-n/2} \int d^n z  \det \left[ {\partial c_i \over
\partial z_j} \right]
\end{equation}
where the factors of $2\pi$ compensate for the absence of $n$ Gaussian
integrations.  To calculate the Jacobian determinant, we first equate
the change in the field resulting from an infinitesimal
change in the $z_i$ with that corresponding to a shift of the $c_i$,
to obtain
\begin{equation}
    \psi_i(x,z) \,\, dc_i =  {\partial \phi_b(x;z) \over \partial z_j}\,\, dz_j 
\end{equation}
Using the orthonormality of the $\psi_j$, we then find that 
\begin{equation}
  (2\pi)^{-n/2} \det \left[ {\partial c_i \over \partial z_j} \right] =
   (2\pi)^{n/2} \det [M_b(z) N^\dagger(z)] = 
          \left[ \det M_b(z) \right]^{1/2}
\end{equation}
so that 
\begin{equation}
   I^{(2)}_b = \int d^nz [\det M_b(z)]^{1/2}
\end{equation}

    The fact that the zero modes all arise from symmetries of the theory might
lead one to expect that the integrand in this equation would be independent of
the $z_j$.  Actually, this is true only if the measure $d^n z$ is invariant
under the symmetry transformations.  If it is not, let $\mu(z)$ be such that 
$\mu(z)^{1/2} d^n z$ is an invariant measure and write
\begin{equation}
   I^{(2)}_b = \int d^nz \mu(z)^{1/2}[\mu(z)^{-1}\det M_b(z)]^{1/2}
\end{equation}
The quantity in brackets is now $z$-independent and can be taken outside the
integral. 

     One can always choose coordinates so that this expression can be written as
a product of a contribution from the translational zero modes and a contribution
from the internal symmetry zero modes.  For the former, the natural choice of
collective coordinates are the spatial coordinates $z^\mu$ of the center of the
bounce. The derivative of the field with respect to these  is, up to a sign, the
same as the spatial derivative of the field. Furthermore, with these coordinates
$\mu(z)=1$, so the integration over the $z^\mu$   simply gives a factor of
$\Omega$.  Hence, the contribution of these modes to $I_b^{(2)}$ is $\Omega
J_b^{\rm trans}$, where\footnote{For the case of a spherically symmetric bounce
in a scalar field theory with a standard kinetic energy term, $J_b^{\rm trans}$
can be expressed in terms of the bounce action, with $J_b^{\rm trans} = 
[S_E(\phi_b) - S_E(\phi_f) ]^2/4\pi^2 $.}
\begin{equation}
    J_b^{\rm trans} = (2\pi)^{-2} \left[\prod_{\mu=1}^4 \int d^4 x
         (\partial_\mu \phi )^2 \right]^{1/2}  
\end{equation}

    The internal symmetry zero modes arise from the action of the
gauge group $G$ on the bounce solution.  Because the bounce tends
asymptotically toward the false vacuum $\phi_f$, normalizable modes
are obtained only from the unbroken symmetry group $H_f \subset G$ of
the false vacuum.  Furthermore, there are no such modes from the
subgroup $K_b \subset H_f$ that leaves the bounce solution invariant.
Hence, the corresponding collective coordinates span the coset space $
H_f/K_b$.  The contribution from these to $I_b^{(2)}$ is then
\begin{equation}
    J_b^{H_f/K_b}\,\, {\cal V}(H_f/K_b)  =  
      [\mu(g_0)^{-1}\det  M_b^{H_f/K_b} (g_0)]^{1/2} \,\, {\cal V}(H_f/K_b)  
\label{Jdef}
\end{equation}  
where ${\cal V}(H_f/K_b) $ is the volume of the coset space, $M_b^{H_f/K_b}
$ is the submatrix corresponding to the internal symmetry zero modes, and
$g_0$ is an arbitrary point of $H_f/K_b$.  To evaluate $J_b^{H_f/K_b}$ it
is convenient to take $g_0$ to correspond to the identity element of
$H_f$.  Writing the group elements near the identity in the form
$e^{i\alpha_jT_j}$, we may take the collective coordinates to be the
parameters that multiply the $T_j$ that that span the coset $H_f/K_b$. 
Evaluated at the identity element, the function $\mu(g)$ is then equal to
unity, while the derivatives with respect to the collective coordinates
are given simply by the action of the generators on the bounce solution. 
Hence \begin{equation}
    J_b^{H_f/K_b} = \left\{\det \left[(2\pi)^{-1} \int d^4x \phi_b^\dagger(x)
           T_j^\dagger T_i \phi_b(x) \right]\right\}^{1/2}
\label{Jresult}
\end{equation}
with the $T_i$ being the generators of $H_f/K_b$. 
Gathering our results together, we obtain
\begin{equation}
    \Gamma =  e^{-B}\,
     {D_b J_b^{\rm trans} J_b^{H_f/K_b} \over D_f}
         \, {\cal V}(H_f/K_b)  
\end{equation}
where $B\equiv S_E(\phi_b) - S_E(\phi_f) $.

It is important to note that $K_b$ is determined
by the symmetry of the bounce, and not by that of the true vacuum; it is
conceivable (although we believe it unlikely) that the latter is invariant
under a larger subgroup $K_t \subset H_f$ than the former.  Even
if $K_t$ is identical to $K_b$, it is not in general the same as the
unbroken symmetry group $H_t \subset G$ of the true vacuum.  For
example, if $G$ is unbroken in the true vacuum and
completely broken in the false vacuum, $H_f$, and hence $K_t$, are
trivial even though $H_t=G$.  In addition, the subgroup $K_b$ depends not
only on the symmetries of the true and false vacua, but also on their
relative orientation.  

   This last point can be illustrated using a theory with global
$SU(5)$ symmetry.  Let us assume that there is a single scalar field
$\phi$, in the adjoint representation, with the potential such that
the false vacuum has unbroken $SU(4)\times U(1)$ symmetry and the
unbroken symmetry of the true vacuum is $SU(3)\times SU(2)\times
U(1)$.  Without loss of generality we may choose the false vacuum
configuration to be of the form
\begin{equation}
  \phi_f = {\rm diag}\, (a,a,a,a,-4a)
\end{equation}
The $SU(5)$ orientation of this field of this configuration influences
that of the true vacuum bubbles that nucleate within it.  Thus, decays
to the true vacua with 
\begin{equation}
    \phi_t^{(1)} = {\rm diag}\, (b,b,b, -{3\over 2} b, -{3\over 2} b) 
\end{equation}
and
\begin{equation}    
   \phi_t^{(2)} = {\rm diag}\, (-{3\over 2} b,-{3\over 2} b, b,b,b )
\end{equation}
are governed by inequivalent bounce solutions and proceed at different
rates \cite{guthew}. If the bounces have the maximum possible symmetry, then in
the former case $K_b = SU(3) \times U(1) \times U(1)$ and there are six
internal symmetry zero modes, while in the latter $K_b = SU(2)\times
SU(2)\times U(1)\times U(1)$ and there are eight such modes.  Of
course, bubble nucleation could with equal probability lead to any
configuration obtained by applying an $SU(4)\times U(1)$
transformation to $\phi_t^{(1)}$ or $\phi_t^{(2)}$; this is taken into
account by the integration over the coset spaces  $(SU(4)\times
U(1))/ (SU(3) \times U(1) \times U(1))$ and $(SU(4)\times 
U(1))/( SU(2)\times SU(2)\times U(1)\times U(1))$.  However, there are true
vacuum configurations that cannot be obtained by such transformations.  In
general, there are no bounce solutions corresponding to these, reflecting the
fact that if a bubble of such a vacuum were to form, the external false vacuum
would exert forces that would realign the field in the bubble interior.

    Finally, let us estimate the magnitude of the zero mode corrections that we
have found.  For definiteness, we will consider the case of a scalar field
theory whose potential can be written as $V(\phi) = \lambda F(\phi)$ with
$F(\phi)$ containing no small dimensionless parameters.  Standard scaling
arguments using the fact that the bounce is a stationary point of the action
show that the bounce has a radius $\sim 1/m$ (where $m$ is a characteristic
scalar mass) and an action (relative to that of the false vacuum) of order
$1/\lambda$.  The typical magnitude of the bounce field is $\phi_b(x) \sim
m/ \sqrt{\lambda} $, while the $T_j$ are all of order unity, so $J_b^{H_f/K_b}
\sim (\lambda m^2)^{-N/2}$, where $N$ is the number of internal symmetry zero
modes.  The ratio $D_b/D_f$ is of order unity in $\lambda$, but is proportional
to a dimensionful parameter $\sim m^{N+4}$ arising from the fact that the
contribution of the zero eigenvalue modes has been deleted from $D_f$. 
Finally, the coset volume is of order unity.  Overall, then, we have
\begin{equation}
   \Gamma =  c_1\, \lambda^{-(N+4)/2} m^4  e^{-c_2/\lambda} 
\end{equation}
where $c_i$ and $c_2$ are of order unity; the effect of the internal symmetry
zero modes has been to enhance the nucleation rate by a factor of order
$\lambda^{-N/2}$.  Phrased somewhat differently, the enhancement is
roughly by a factor of $B^{N/2}$.

\section{$SU(2)$ Symmetry}

      As a concrete example, let us consider the case where the symmetry
group of the false vacuum is $H_f=SU(2)$ but the bounce solutions break this
symmetry.  A natural set of collective coordinates is given by the Euler
angles.  Thus, given one bounce solution $\phi_b^0(x)$, we can define a
three-parameter family of solutions by
\begin{equation}
    \phi_b(x;\varphi,\theta,\psi) = e^{i\varphi T_3}\, e^{i\theta T_2}
    \, e^{i\psi T_3} \, \phi_b^0(x)  \equiv U(\varphi,\theta,\psi)  \phi_b^0(x) 
\end{equation}
where the $T_j$ are the appropriate (possibly reducible) representation of the
generators of $SU(2)$.  Differentiation of this expression gives
\begin{eqnarray}
    \partial_\varphi  \phi_b(x;\varphi,\theta,\psi) &=& iU(\varphi,\theta,\psi)  
        \tilde T_3 \phi_b^0(x)   \nonumber \\
     \partial_\theta  \phi_b(x;\varphi,\theta,\psi) &=&i U(\varphi,\theta,\psi)  
        \tilde T_2 \phi_b^0(x)  \nonumber \\
    \partial_\psi  \phi_b(x;\varphi,\theta,\psi) &=& iU(\varphi,\theta,\psi)  
       T_3  \phi_b^0(x) 
\end{eqnarray}
where
\begin{eqnarray}
     \tilde T_3 &=& \, e^{-i\psi T_3} \,e^{-i\theta T_2}\, T_3\,
       e^{i\theta T_2}\, e^{i\psi T_3} \nonumber \\
           &=&  \cos\psi \sin\theta \,T_1 + \sin\psi \sin\theta \,T_2 
               + \cos\theta\, T_3   \nonumber \\
     \tilde T_2 &=& \, e^{-i\psi T_3} \, T_2\, e^{i\psi T_3} \nonumber \\
           &=& -\sin\psi\, T_1  + \cos\psi\, T_2  
\end{eqnarray}
Thus, if $z_j=(\varphi,\theta,\psi)$,
\begin{equation}
      \partial_j \phi_b(x;\varphi,\theta,\psi) 
         = i K_{jk} \,U(\varphi,\theta,\psi) T_k \phi_b^0(x) 
\end{equation}
where
\begin{equation}
       K = \left( \matrix{\cos\psi \sin\theta &   \sin\psi \sin\theta & 
       \cos\psi\cos\theta  \cr
        -\sin\psi   &   \cos\psi & 0 \cr  0 & 0 & 1 } \right) 
\end{equation}
Substitution of this into Eq.~(\ref{Mdef}) yields
\begin{equation}
     (M_b^{SU(2)/K_b})_{il}  =(2\pi)^{-1} K_{ij} K_{kl}  \int d^4x
      \phi_b^{0\dagger}(x)T_j^\dagger T_k \phi_b^0(x) 
\end{equation}
Now recall that an invariant measure on $SU(2)$ is given by $\sin\theta
d\varphi \,d\theta \, d\psi$, so we may take $\mu(\varphi,\theta,\psi)
=\sin^2\theta = (\det K)^2$.  Hence,
\begin{equation}
     \mu^{-1} \det M_b^{SU(2)/K_b}  = \det \left[ (2\pi)^{-1}  \int d^4x
      \phi_b^{0\dagger}(x)T_j^\dagger T_k \phi_b^0(x) \right]
\label{Mresult}
\end{equation}
This is independent of the collective coordinates, as promised, and is in
agreement with Eq.~(\ref{Jresult}).

Three specific cases may serve to illustrate some of the possible behaviors:

a) One $SU(2)$ doublet:  If the bounce involves an $SU(2)$ doublet, then 
the bounce completely breaks the $SU(2)$ symmetry.   The coset volume factor is
\begin{equation}
    {\cal V}(H_f/K_b)  =  {\cal V}(SU(2)) = 16\pi^2
\end{equation}
while $M_b^{SU(2)/K_b} = M_b^{SU(2)}$ is proportional to the unit matrix, with
\begin{equation}
     \mu^{-1} \left[M_b^{SU(2)}\right]_{ij} = \delta_{ij}  
     (2\pi)^{-1}  \int d^4x \phi_b^\dagger(x)  \phi_b(x) 
\end{equation}
(Because our formulas have been derived using real fields, one must use a
four-dimensional real representation, rather than a two-dimensional complex
representation, for the $T_j$ when obtaining this result from
Eq.~(\ref{Mresult}).)

b) One $SU(2)$ triplet:  If the bounce is constructed from a single real
triplet whose direction is independent of $x$ [i.e., such that
$\phi_b(x)$ can be written as $(0,0,f(x))$], then $K_b=U(1)$.  There are only
two zero modes and $H_f/K_b$ is the two-sphere spanned by $\theta$ and
$\varphi$ with 
\begin{equation}
    {\cal V}(H_f/K_b)  =  {\cal V}(SU(2)/U(1)) = 4\pi
\end{equation}
$M_b^{SU(2)/K_b}$ is now a $2\times 2$ matrix, with 
\begin{equation}
     \mu^{-1} \left[M_b^{SU(2)/U(1)}\right]_{ij} = \delta_{ij}  
     (2\pi)^{-1}  \int d^4x {\mbox{\boldmath $\phi$}}^2_b(x) 
\end{equation}

c) Two non-parallel $SU(2)$ triplets:  If the bounce solution contains two
triplet fields that are not parallel, then the bounce has no continuous
symmetry.  Because only integer spin fields are involved,  \begin{equation}
    {\cal V}(H_f/K_b)  =  {\cal V}(SO(3)) = 8\pi^2
\end{equation}
The matrix $\mu^{-1} M_b^{SO(3)}$ has three unequal eigenvalues.

\section{Bubble Nucleation in Curved Space-Time}

    Coleman and De Luccia \cite{cdl} showed that the bounce formalism could
be extended to include the effects of gravity by requiring that both the
bounce and the homogeneous false vacuum configurations be solutions of the
coupled Euclidean matter and Einstein equations.   For a scalar field theory
with $V(\phi) \ge 0$, as we henceforth assume, the false vacuum solution
consists of a uniform scalar field $\phi_f$ on a four-sphere of
radius  
\begin{equation}
  \tilde H_f^{-1} = \sqrt{ 3 M_{\rm Pl}^2 \over 8\pi V(\phi_f)}
\end{equation}
with total Euclidean action (including gravitational contributions) 
\begin{equation}
 S_E(\phi_f) = -{3 M_{\rm Pl}^4 \over 8 V(\phi_f)}
\end{equation}
The bounce solution has the same topology, with regions of
approximate true vacuum and false vacuum separated by a wall region.  If
the matter mass scale $\cal M$ is much less than the Planck mass $M_{\rm Pl}$,
then both the radius $R_b$ of the true vacuum region and the difference between
the bounce action and the false vacuum action differ from the corresponding
flat space quantities by terms of order $({\cal M}/M_{\rm Pl})^2$.

    The spectra of the small fluctuations about these solutions again
contain one zero mode for each symmetry of the Lagrangian that is broken by
the solution.  However, because the Euclidean 
solutions are on closed manifolds with finite volumes, the modes due
to symmetries broken by the false vacuum are normalizable, in contrast with the
flat space case.  Hence, we would expect the 
flat space factors given in Eq.~(\ref{Jdef}) to be replaced by 
\begin{equation}
    { J_b^{G/K_b}  \over  J_f^{G/H_f} } \,
       {{\cal V}(G/K_b) \over  {\cal V}(G/H_f) }
      =  { J_b^{G/K_b}  \over  J_f^{G/H_f} }
          \,  {\cal V}(H_f/K_b)
\label{curvedJac}
\end{equation}
Although the volume factors give the same result as in flat space, the Jacobean
factors appear quite different.  Yet, for ${\cal M} \ll M_{\rm Pl}$, where
gravitational corrections should be small, this should approach the flat space
result.  

      To see how this comes about, let us denote by $t_j$ the generators of
$H_f/K_b$ and by $s_j$ those of $G/H_f$.  The Jacobean determinant in the numerator
of Eq~(\ref{curvedJac}) has contributions from matrix elements containing both
types of generators, whereas the determinant in the denominator only involves
$s_i s_j$ matrix elements.   Because the $t_j$ annihilate the false vacuum, the
matrix elements involving these have nonzero contributions only from the
region, of volume $\sim R_b^4$, where the bounce solution differs from the
false vacuum and hence are suppressed by a factor of order $(\tilde
H_fR_b)^4 \sim 
({\cal M}/M_{\rm Pl})^4$ relative to the $s_i s_j$ matrix elements.  
(We are assuming that $R_b\sim {\cal M}^{-1}$; this will be the case for
generic values of the parameters.) This implies
that, up to corrections of order $({\cal M}/M_{\rm Pl})^8$, the determinant can
be written as a product of a determinant involving only the $t_i$ and one
involving only the $s_i$; i.e., 
\begin{equation}
    { J_b^{G/K_b}  \over  J_f^{G/H_f}} = { J_b^{G/H_f}  \over  J_f^{G/H_f} }\,
             J_b^{H_f/K_b} \, [ 1+ O(({\cal M}/M_{\rm Pl})^8) ]
\end{equation}
The first factor on the right hand side differs from unity only by an amount
proportional to the fraction $\sim ({\cal M}/M_{\rm Pl})^4$ of the Euclidean
space where the bounce differs from the false vacuum.  The second factor differs
from the corresponding flat-space term only by the replacement of the
matter fields of the flat space bounce by those of the curved space bounce, and
so clearly reduces to the flat space result as ${\cal M}/M_{\rm Pl}\rightarrow
0$.

     The fact that the bounce solution is a closed manifold, with the
true and false vacuum regions both finite, suggests that it
can contribute not only to the nucleation of a true vacuum bubble within
a false vacuum background, but also to the nucleation of a false vacuum
bubble within a true vacuum background, with the rate for the latter
process obtained from that of the former by making the substitution
$\phi_f \rightarrow \phi_t$ \cite{truedecay}.  To leading order, the ratio of
these two rates is  \begin{equation} {\Gamma_{t\rightarrow f} \over
\Gamma_{f\rightarrow t}} 
     =   e^{S_E(\phi_t) -S_E(\phi_f)} 
    = \exp\left[ -{3M_{\rm Pl}^4 \over 8}\left({1\over V(\phi_t)} -
      {1\over V(\phi_f)}\right) \right]
\end{equation} 

     The continued nucleation and expansion of bubbles of one vacuum within
the other will result in a spacetime that is a rather inhomogeneous
mixture of the two vacua.   There is an intriguing thermal 
interpretation of this mixture if $V(\phi_f) - V(\phi_t)
\ll (V(\phi_f) + V(\phi_t))/2 \equiv \bar V$, so that the geometry of space is
approximately the same in the regions of either vacua, with a Hubble parameter
\begin{equation}  
    \bar H \approx  \sqrt{8\pi \bar V \over 3 M_{\rm Pl}^2}    
\end{equation}

It seems plausible that the fraction of space contained in each of the
vacua might tend to a constant, with the nucleation of
true vacuum bubbles in false vacuum regions being just balanced by the
nucleation of false vacuum bubbles in true vacuum regions.  For such an
equilibrium to hold, the volumes $\Omega_f$ and $\Omega_t$ of false
and true vacuum must satisfy
\begin{equation} { \Omega_f \over \Omega_t} = 
   {\Gamma_{t\rightarrow f} \over \Gamma_{f\rightarrow t}} \approx
      e^{-\Omega_{\rm hor} [V(\phi_f) - V(\phi_t)]/T_H }
\label{rateratio}
\end{equation}
where the horizon volume $\Omega_{\rm hor}= (4\pi/3) \bar H^{-3}$ and the
Hawking temperature $T_H = \bar H/2\pi$.  If we view the de Sitter space as
being somewhat analogous to an ensemble of quasi-independent horizon volumes in
a thermal bath, then this leading contribution to the volume ratio is
essentially a Boltzmann factor.  

    The zero mode corrections to the nucleation rate are consistent with this
thermodynamic picture.  Their effect is to multiply the ratio in
Eq.~(\ref{rateratio}) by    
\begin{equation}
    {{\cal V}(G/H_f) \over {\cal V}(G/H_t)} \, {J_f^{G/H_f} \over
J_t^{G/H_t}  } 
     =  \left ( {\bar H\over \sqrt{3}\,\pi} \right)^{N_t-N_f} \,
    {{\cal V}(G/H_f) \over {\cal V}(G/H_t)}\,
    \left[{ \det\left[ (\Omega_{\rm hor}T_H /2\pi) \left(\phi_f^\dagger T_i T_j
\phi_f\right)     \right]
      \over \det\left[ (\Omega_{\rm hor}T_H/2\pi )\left(\phi_t^\dagger T_i
  T_j  \phi_t \right)\right] }\right]^{1/2}  
\label{ratioeq}
\end{equation}
where $N_f$ and $N_t$ are the number of internal symmetry zero modes in the false
and true vacua, respectively.  We recognize the dimensionless ratio on the
right hand side as the ratio of two classical partition functions of the
form 
\begin{equation}
       \int {d^Nz d^Np\over (2\pi)^N} e^{-{\cal H}_z/T_H}  = 
      \int d^N z \left[ (\Omega_{\rm hor}T_H /2\pi)
    \left(\phi^\dagger T_i T_j \phi \right)\right]^{1/2} 
\end{equation}
that follow from the effective Lagrangian
\begin{equation}
    L_z = {1\over 2} \Omega_{\rm hor} \left(\phi^\dagger T_i T_j \phi
    \right) \,  \dot z_i \dot z_j
\end{equation} 
that describes the collective coordinates dynamics for a horizon volume
with spatially uniform scalar field $\phi$. 

 The presence of a dimensionful prefactor in Eq.~(\ref{ratioeq}) is required
by the differing numbers of zero modes about the true and false vacua, which
implies a dimensional mismatch between the functional determinants over the
nonzero eigenvalue modes.  This suggests that the factor of $(\bar
H/\sqrt{3}\, \pi)^{N_t-N_f}$ should, like the functional determinants
themselves, be 
understood as related to the first quantum corrections to the vacuum
energies.

\section{Concluding Remarks} 

We have seen that when a metastable false vacuum decays to a true
vacuum that breaks some of the internal symmetries of the false
vacuum, the presence of $N>0$ zero modes about the bounce
solution can lead to an enhancement of the bubble nucleation rate.
In a theory characterized by an overall scalar coupling $\lambda$,
the zero-temperature, quantum mechanical, tunneling
rate is increased by a factor of order $\lambda^{-N/2} \sim B^{N/2}$.  
A straightforward extension of our methods to finite temperature
thermal tunneling shows that, although the $\lambda$-dependence is
changed, the enhancement is still of order $B^{N/2}$.  Since the
nucleation rate falls exponentially with $B$, we therefore have
the curious situation that the enhancement is greatest when the
overall rate is smallest.  

These results may be of particular interest for the symmetry-breaking
phase transitions that arise in the context of a grand unified
theory.  For any given nucleation rate, the numerical effect of the
zero mode corrections will, of course, almost always be negligible
compared to the uncertainties due to the undetermined couplings in the
scalar field potential.  The zero mode effects could, however, be
significant when there are competing decays to vacua with different
degrees of symmetry breaking, such as are encountered in many
supersymmetric models.

\vskip 1cm
\centerline{\large\bf Acknowledgments}
\vskip  5mm

A.K and E.W. would like to thank the Aspen Center for Physics, where
part of this work was done.  This work was supported in part by the
U.S. Department of Energy. K.L. is supported in part by the NSF
Presidential Young Investigator program.

\end{document}